\documentclass[pra,aps,twocolumn,showpacs]{revtex4-1}

\usepackage[english]{babel}
\usepackage{multirow}
\usepackage{pifont}                           
\usepackage{amssymb}                           
\usepackage{amsmath}                           
\usepackage{graphicx}                          
\usepackage{bm}                                
\usepackage{color}
\usepackage{psfrag}
\usepackage{fix-cm}

\begin{document}

\title{{\it Ab initio} methods for finite temperature two-dimensional Bose gases}

\author{S.\ P.\ Cockburn}
\author{N.\ P.\ Proukakis}

\affiliation{Joint Quantum Centre (JQC) Durham-Newcastle, 
School of Mathematics and Statistics,\\ Newcastle University, 
Newcastle upon Tyne, NE1 7RU, United Kingdom
}

\date{\today}


\begin{abstract}

The stochastic Gross-Pitaevskii equation and modified Popov theory are shown to provide an {\it ab initio} 
description of finite temperature, weakly-interacting two-dimensional Bose gas experiments. 
Using modified Popov theory, a systematic approach is developed in which the 
momentum cut-off inherent to classical field methods is removed as a free parameter. 
This is shown to yield excellent agreement with the recent experiment of Hung {\it et al.} 
[Nature, {\bf 470} 236 (2011)], verifying that the stochastic Gross-Pitaevskii equation 
captures the observed universality and scale-invariance.

\end{abstract}

\maketitle


\textit{Introduction --- }
Ultracold atomic gases represent versatile tools with which to investigate many-body quantum physics
\cite{Bloch2008}. 
A key tunable property of these systems is their effective dimensionality; increasing the trapping 
potential, used in experiments to confine and cool atoms, in one or two directions produces gases 
that are effectively two-dimensional (2D) or one-dimensional (1D), respectively.
Low-dimensional geometries in turn lead to richer physics due to 
the enhanced importance of fluctuations which restrict the onset of long range order 
\cite{Mermin1966,Hohenberg1967}.

A number of experiments have been performed recently in order to 
examine the thermodynamic properties of weakly interacting 2D Bose gases 
\cite{Hadzibabic2006,Schweikhard2007,Kruger2007,Hadzibabic2008,Clade2009,Tung2010,Rath2010,Yefsah2011,
Plisson2011,Hung2011}.
Due to the high precision now routinely attained in such experiments, a particularly powerful 
feature is their usefulness in accurately testing microscopic theories.
In this respect, the 2D Bose gas is interesting as 
fluctuations are typically important over a broad critical region \cite{Prokofev2002,Posazhennikova2006},
meaning the standard mean field approach to weakly interacting Bose gases is not well suited;
this broadness was however exploited by Hung {\it et al.} \cite{Hung2011} to obtain
a clear experimental observation of critical phenomena in ultracold atoms near the Berezinskii-Kosterlitz-Thouless (BKT) phase 
transition \cite{Berezinskii1972,Kosterlitz1973}.

At equilibrium, Monte Carlo (MC) calculations have been 
successfully applied to the uniform 2D Bose gas \cite{Kagan2000,Svistunov2001,Prokofev2002}, 
notably yielding a microscopic prediction for the BKT transition point \cite{Svistunov2001}. 
The harmonically trapped case has been studied using quantum MC \cite{Holzmann2008}, 
various mean field theories \cite{Gies2004a,Gies2004b,LihKing2008,Holzmann2008b,Bisset2009c}
and classical field calculations \cite{Simula2008,Bisset2009a,Bisset2009b}.
Classical field methods have the advantage of providing 
a time-dependent description of the gas, 
which makes them additionally applicable to systems away from equilibrium;
nonetheless, a simple yet accurate mean-field theory is also highly desirable
to avoid the need for more complicated methods in calculating equilibrium properties.
In this work we perform 
a quantitative comparison between 
two such complementary methods:
(i) the modified Popov (MP) theory of Stoof and co-workers 
\cite{Andersen2002,AlKhawaja2002}, 
which is straightforward to solve
and provides quick access to equilibrium properties in all dimensions, 
and (ii) the stochastic Gross-Pitaevskii equation (SGPE) \cite{Stoof1999,Stoof2001,Duine2001} 
(see also \cite{Gardiner2003})  
that has already successfully described several 1D Bose gas experiments \cite{Cockburn2011a,Gallucci2012}
and is also applicable in dynamical situations \cite{Stoof2001,Cockburn2010}.
The experimental {\it in situ} measurements of Hung {\it et al.}~\cite{Hung2011} 
offer an ideal test-bed for the theories we wish to consider, free 
of the complications associated with modelling expansion imaging,
thus a primary motivation for this work is to demonstrate the applicability
of the SGPE in capturing the universality and scale-invariance of 2D experiments.

A further related aim is to systematically circumvent a
standard limitation of classical field calculations,
linked to the fact that the Bose gas represented in this way must obey classical statistics.
In practical terms, this issue manifests as a sensitivity to the numerical grid 
(or, equivalently, momentum/energy cutoff) used in solving such models \cite{Cockburn2009,Blakie2008},
which, as a free parameter, can undermine their use for {\it ab initio} studies.
While calculations based upon classical lattice models have been linked back to their quantum 
counterparts in the context of Monte Carlo studies \cite{Svistunov2001,Arnold2001}, 
the situation is less clear for methods based upon {\it dynamical} equilibration of 
classical fields: approaches to date include 
selecting a cutoff based on the assumption that the classical field temperature should 
match that of an {\it ideal} Bose gas with the same condensate fraction \cite{Zawitkowski2004},
specifying a minimum acceptable mode occupation, typically within an 
energy cutoff somewhere in the range $g_{\rm 2D}n<E_{\rm cut}\lesssim k_{B}T$ \cite{Blakie2008}, and
a high temperature semi-classical field method valid for $k_{B}T>\mu$ \cite{Giorgetti2007}.
Motivated by the desire for a formulaic approach to cutoff choice that is valid also at 
temperatures $k_{B}T\lesssim\mu$, and which also includes the effect of interactions, in 
this work we have devised and tested a strategy by which to specify an `optimum' SGPE 
grid choice, based upon comparing the classical and quantum limits of MP theory.

\textit{ Methodology --- }
The 2D stochastic Gross-Pitaevskii equation describes the Bose gas via a noisy 
field $\psi({\bf x},t)$, which satisfies the equation of motion
\begin{equation}
  \begin{split}
    i\hbar&\frac{\partial \psi({\bf x},t)}{\partial t} = (1-i\gamma({\bf x},t))
    \bigg[-\frac{\hbar^{2}}{2m}\nabla^{2}_{\rm x,y}+V({\bf x})-\mu\\
    &+g_{\rm 2D}\left(|\psi({\bf x},t)|^{2}+2n_{\rm above}({\bf x})\right)\bigg]\psi({\bf x},t)+\eta({\bf x},t),
  \end{split}
  \label{eq:SGPE}
\end{equation}
where $V({\bf x})=m\omega^{2}(x^{2}+y^{2})/2=m\omega^{2}r^{2}/2$ is the trapping potential in the more 
weakly confined x-y plane, $g_{\rm 2D}=\sqrt{8\pi}(a/l_z)\hbar^2/m=g\hbar^2/m$
is the 2D interaction strength (with $a$ the s-wave scattering length),
and $\eta$ is a complex Gaussian noise term, with correlations given by the relation
$\langle \eta^{*}({\bf x},t) \eta({\bf x}',t') \rangle =2\hbar\gamma({\bf x},t)k_{B}T\delta({\bf x}-{\bf x}')\delta(t-t')$. 
In Eq.~\eqref{eq:SGPE} $n_{\rm above}$ denotes the density of atoms with momenta greater than the momentum
cutoff due to the numerical grid used to solve the SGPE 
\footnote{The density of atoms with momenta greater than the grid cutoff, $n_{\rm above}$,
is calculated iteratively, in the Hartee-Fock limit, prior to carrying out the SGPE simulations
using the density from MP theory based upon Rayleigh-Jeans statistics.}.

Classical field methods arise following the observation that the Bose field 
operator may be accurately replaced in the Heisenberg equation of motion
by a complex valued field, under the condition that system modes represented in this 
way are highly occupied. Although this is an excellent approximation in many circumstances, 
the system is then found to obey Rayleigh-Jeans statistics at equilibrium.
In arriving at Eq.\eqref{eq:SGPE}, a similar notion is embodied in 
moving to a classical fluctuation-dissipation theorem
(see Eqs.(38)-(40) of Ref.~\cite{Duine2001}).

The numerical solution to Eq.\eqref{eq:SGPE} introduces an ultraviolet momentum cutoff 
implemented here by the discretization scheme chosen;
this implies that the equilibrium thermal state that is achieved for a given set of physical 
parameters can differ quite dramatically through variation of the grid spacing alone. 
There is then clearly some ambiguity as to which cutoff choice provides optimum agreement with nature. 
\begin{figure}[t!]
  \centering
  \includegraphics[angle=0,scale=0.30,clip]{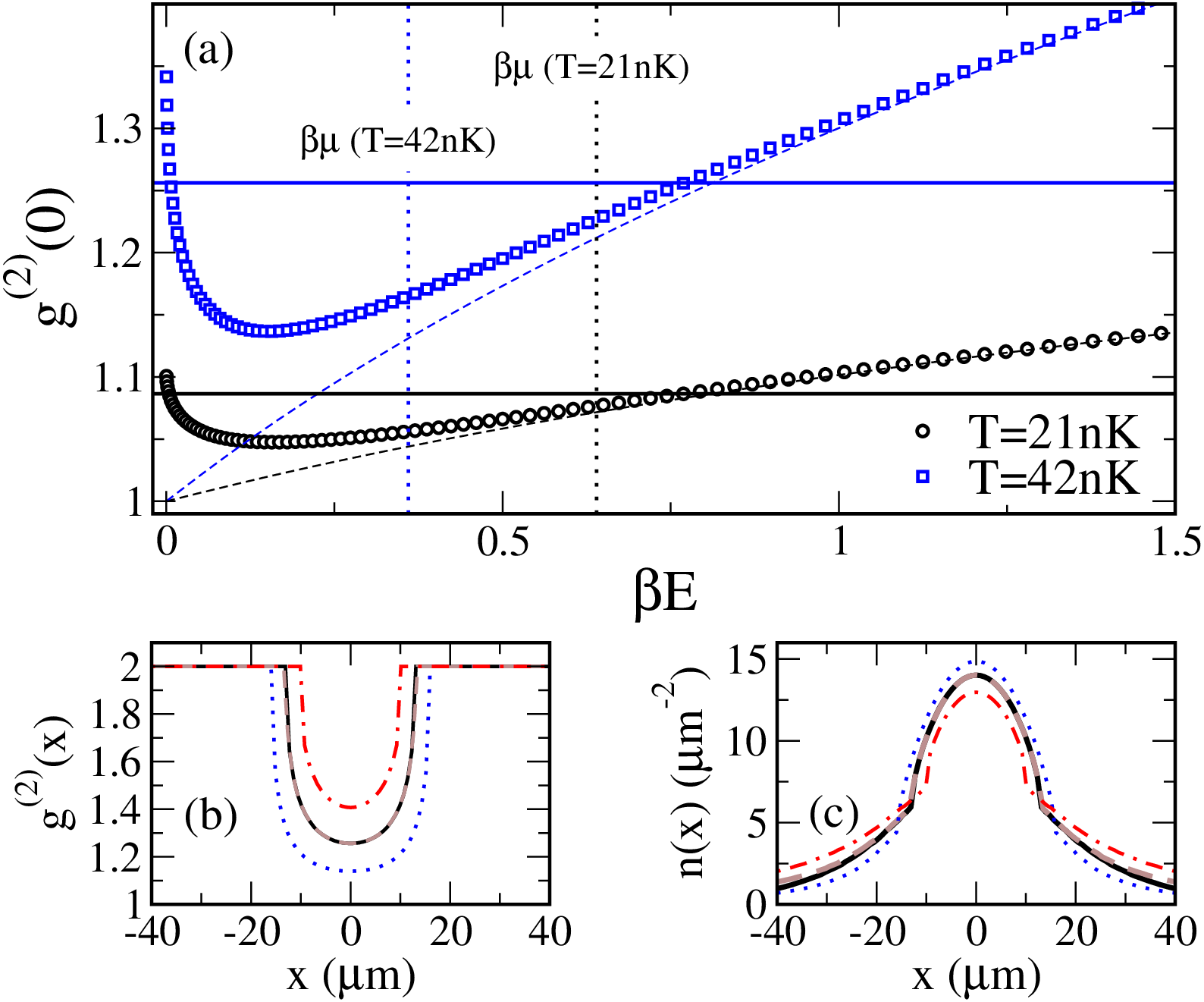}
  \caption {Optimum grid choice: (a) 
	Variation in $g^{(2)}(0)$ with energy cutoff 
  for MP$_{\rm RJ}$
  at two different temperatures. 
  Horizontal lines indicate the MP$_{\rm BE }$ results;
	the horizontal coordinate of the point at which the MP$_{\rm RJ}$ data intersects the relevant
  MP$_{\rm BE }$ result identifies the optimum energy cutoff.
	Dashed lines indicate the results when above cutoff atoms are neglected. 
	(b) $g^{(2)}(x)$ 
	and (c) density profiles from MP$_{\rm BE}$ 
  (solid black) and MP$_{\rm RJ}$ for
  several cutoffs: $\beta E=0.20$ (dotted blue),
  $\approx0.77$ (dashed brown) 
  and $=1.50$ (dot-dashed red).
  }
  \label{fig:opt_grid}
\end{figure}
To address this point for 2D Bose gases, we choose to make use of the MP theory.

The MP theory \cite{Andersen2002,AlKhawaja2002}
was formulated specifically to describe low-dimensional Bose gases, 
taking into account the effects of phase fluctuations to all orders \cite{Andersen2002}, and 
has been found to agree well with the SGPE in previous studies \cite{AlKhawaja2002,Proukakis2006b,Cockburn2011b}.
In Ref.\cite{LihKing2008}, MP was compared to both Hartree-Fock (HF) and MC results for the 2D Bose gas.
The MP predictions for the critical chemical potential and density for the BKT 
transition were in excellent agreement with those obtained from MC simulations \cite{Svistunov2001}.
In fact, the MP theory was found only to break down in the region
where its equation of state reduces to that of HF,
which occurs close to the BKT transition
due to the mean field nature of this approach. A renormalization 
group analysis was subsequently shown to avoid this discontinuity and to match smoothly 
to the MP equation of state for chemical potentials $\mu>0.18k_{B}T$ \cite{LihKing2008}.

\begin{figure*}[ht!]
  \centering
  \includegraphics[angle=0,scale=0.34,clip]{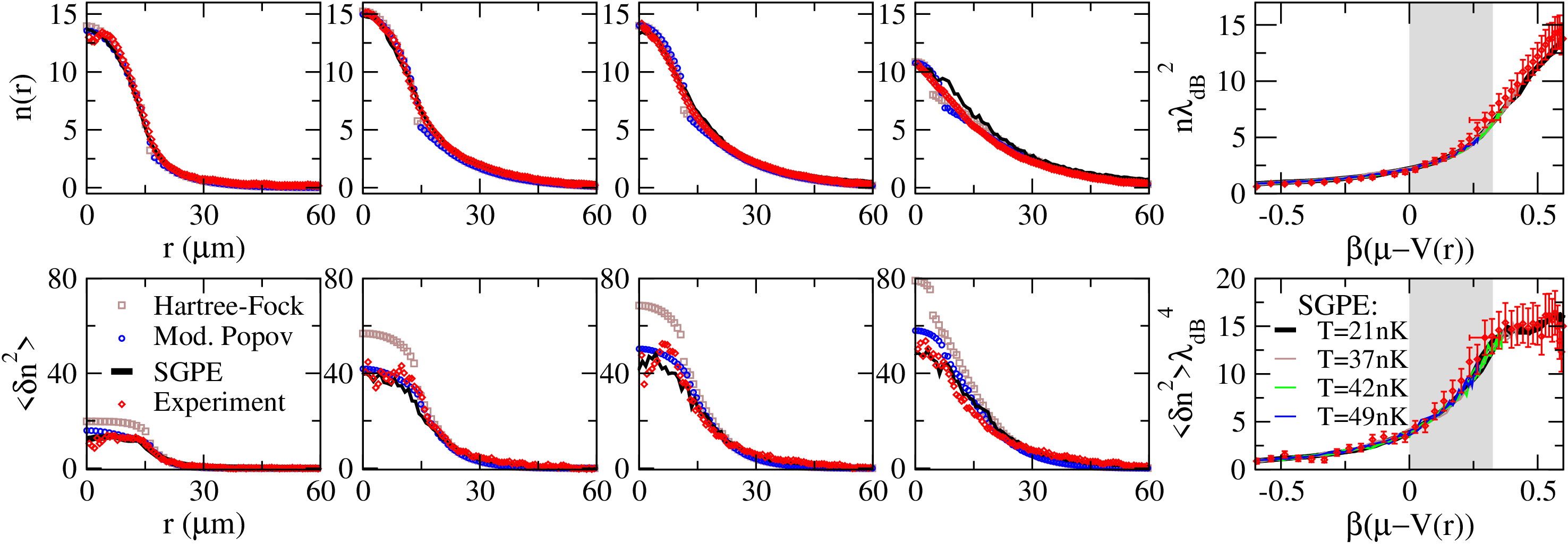}%
  \caption {Density (top) and density fluctuations (bottom) from the SGPE, modified Popov (Bose-Einstein) 
	and Hartree-Fock calculations versus the experimental	data of Ref.~\cite{Hung2011}. 
  The rightmost plots show the SGPE results capture the experimentally observed scale invariance found in
	Ref.~\cite{Hung2011}, including across the transition region indicated by the grey shaded area.
  }
  \label{dens}
\end{figure*} 
The central object within the MP approach is the quasi-condensate \cite{PopovBook}
which, in the Thomas-Fermi approximation, obeys 
$g_{\rm 2D}(n_{\rm qc}+2n_{\rm t})=\mu_{\rm eff}$
where the effective chemical potential may be written as 
$\mu_{\rm eff}=\mu-V({\bf x})$. The thermal density is calculated from \cite{Andersen2002,AlKhawaja2002}
\begin{equation} 
  n_{\rm t}=n_{\rm qc}+\frac{1}{V}\sum_{{\bf k}=0}^{{\bf k_{\rm max}}}
  \left\{\frac{\epsilon_{{\bf k}}}{2\hbar\omega_{{\bf k}}}\left[2N_{{\bf k}}+1\right]-
  \frac{1}{2}+\frac{g_{\rm 2D}n_{\rm qc}}{2\epsilon_{{\bf k}}+2\mu}\right\},
	\label{eq:mP}
\end{equation} 
with $\hbar\omega_{{\bf k}}=\sqrt{\epsilon_{{\bf k}}(\epsilon_{{\bf k}}+2g_{\rm 2D}n_{\rm qc})}$ and $\epsilon_{{\bf k}}=\hbar^{2} {\bf k}^{2}/2m$.

\textit{Optimum cutoff selction: Classical vs. quantum statistics --- }
In the usual formulation of MP, 
$N_{{\bf k}}=N^{\rm BE}_{{\bf k}}\equiv1/(\exp(\beta\hbar\omega_{{\bf k}})-1)$, since the particles 
obey Bose-Einstein statistics, and the upper index 
in the sum in Eq.~\eqref{eq:mP}, ${\bf k_{\rm max}}$, may be straightforwardly taken to infinity. 
However, setting $N_{{\bf k}}=N^{\rm RJ}_{{\bf k}}\equiv 1/(\beta\hbar\omega_{{\bf k}})$ 
instead leads to a MP result based upon Rayleigh-Jeans statistics
which is therefore analogous to using the classical fluctuation-dissipation 
result in the SGPE (see e.g. Eqs.~(38)-(40) of Ref.~\cite{Duine2001}).
We will use MP$_{\rm RJ}$ and MP$_{\rm BE}$ to denote the Rayleigh-Jeans and 
Bose-Einstein cases, respectively.

Since the MP$_{\rm RJ}$ approach gives very good agreement with the SGPE 
for a given cutoff 
\footnote{As in the SGPE, we also include the atoms with momenta above the cutoff 
in the classical case of MP. For the optimum cutoff, we find this number
of atoms to be a small fraction of the total, which we retain mainly for completeness.},
this allows us to extract an optimum value for the SGPE grid spacing,
as we can examine the accuracy of the classical approximation
as ${\bf k_{\rm max}}$ is varied by directly comparing the 
MP$_{\rm RJ}$ and MP$_{\rm BE}$ results. 
To measure this difference, we compare predictions for the normalized second-order correlation function, 
chosen because of the important role played by density fluctuations in quasicondensation.
In MP theory, this is given by $g^{(2)}=(n_{\rm qc}^2+4n_{\rm qc}n_{\rm t}+2n_{\rm t}^2)/(n_{\rm qc}+n_{\rm t})^2$.
The procedure we adopt then is to first calculate $g^{(2)}({\bf x}=0)$ using the 
Bose-Einstein distribution in Eq.~\eqref{eq:mP}. 
This value then serves as a target result which we aim to match using MP$_{\rm RJ}$
with all other parameters held constant. Once an optimum cutoff is selected
in this way, it is then used to calculate the optimum numerical grid spacing for use
in the SGPE simulations.

This approach is illustrated in Fig.~\ref{fig:opt_grid}, 
symbols indicating the $g^{(2)}(0)$ values obtained 
from MP$_{\rm BE}$ for two temperatures as the energy cutoff, shown on the horizontal axis, is varied.
The diagonal dashed lines show the results if we neglect atoms above the cutoff
momentum in the MP$_{\rm RJ}$ calculations. The symbols asymptote towards these lines for energy
cutoffs~$\gtrsim k_{B}T$, as 
there are then very few atoms in modes above the cutoff energy.
For each temperature, the optimum energy cutoff may be read off from Fig.~\ref{fig:opt_grid}
as the horizontal coordinate of the point at which the classical $g^{(2)}(0)$ data
(squares/circles) 
intersects the horizontal line indicating the Bose-Einstein result.
The lower plots show $g^{(2)}(x)$ and density profiles for the Bose-Einstein
calculation (solid black) and the Rayleigh-Jeans for several cutoffs: $\beta E=0.20$ (dotted blue),
$0.77$ (dashed brown) and $1.50$ (dot-dashed red).

\begin{figure}[b!]
  \centering
  \includegraphics[angle=0,scale=0.3,clip]{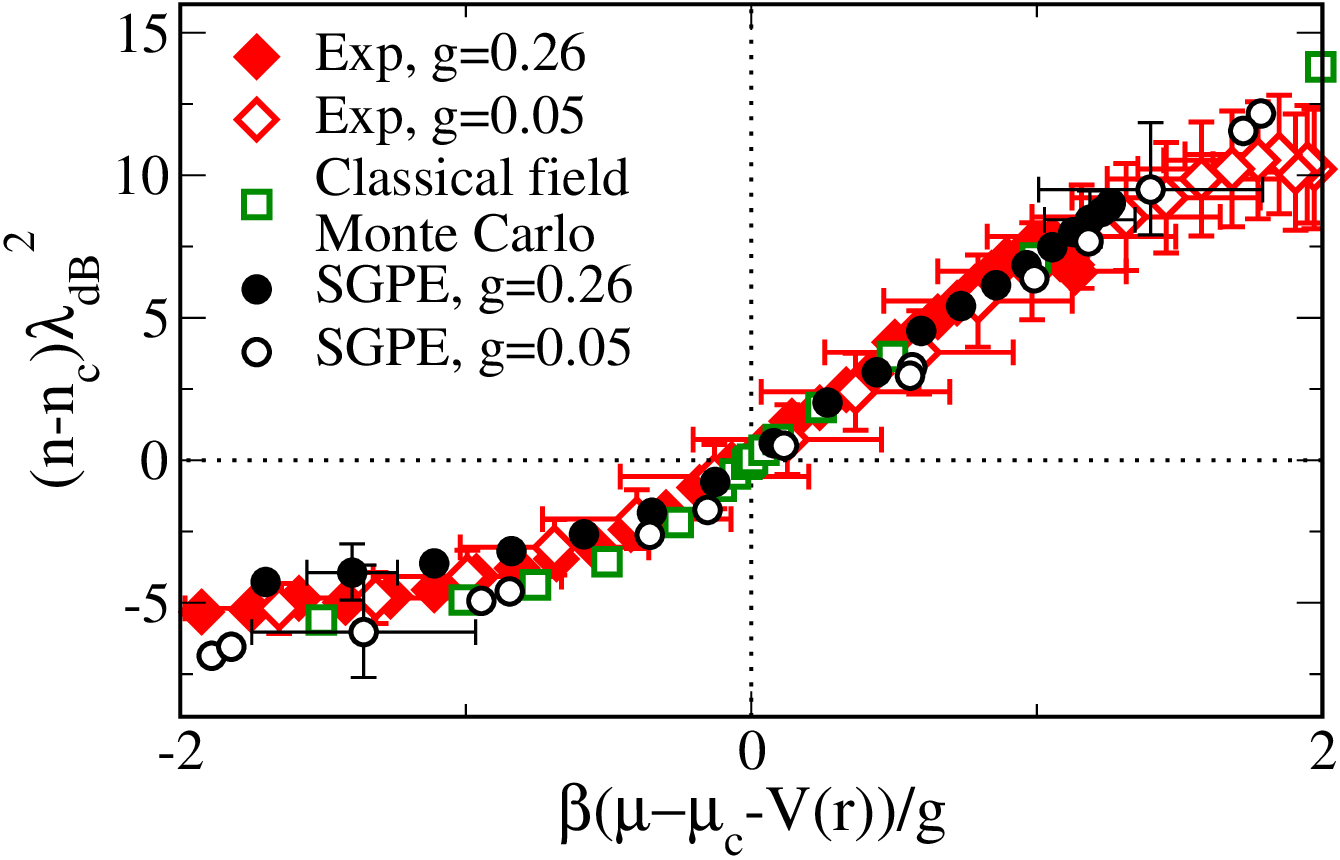}
  \caption{Universal behaviour of shifted 
	density profiles $(n-n_{\rm c})\lambda_{\rm dB}^{2}$ versus 
	rescaled chemical potential $\beta(\mu-\mu_{\rm c}-V(r))/g$. Indicative errorbars are shown for several SGPE 
	data points which originate from the fit used to obtain $\mu_{\rm c}$.}
	\label{crit}
\end{figure}
For the {\it ideal} gas, an expression was derived in Ref.~\cite{Witkowska2009}
for a cutoff leading to optimum agreement between quantum and classical
statistics; 
in 2D, it was found that this occured at an energy $\beta E\approx1.6$.
Using the method outlined above we typically obtain a value around 
$\beta E\approx0.77$, which varies slowly with temperature in the range 
considered, yet is clearly much lower than the ideal gas prediction,
indicating the important role of interactions.

\textit{ Comparison to Experimental results --- }
In order to validate our scheme, we now directly compare the results 
to {\it in situ} experimental data of Hung {\it et al.} \cite{Hung2011}. 
We begin with the MP$_{BE}$ method
and fix the trap parameters and temperature to those quoted in \cite{Hung2011}. 
We then vary the chemical potential until the required density is reached at the trap centre. 
This provides us with all relevant physical parameters required 
to solve the SGPE, barring an optimum choice of grid spacing, which is extracted
as described; since there are no free parameters, both the MP theory and SGPE provide  
an {\it ab initio} description of the experiment.

Figure~\ref{dens} shows the SGPE (with an optimum momentum cutoff) and MP
results together with the experimental density profiles of Ref.~\cite{Hung2011}.
Corresponding data for density fluctuations, 
obtained from the normalized second order correlation function $g^{(2)}(r)$ via
$\langle\delta n^{2}(r)\rangle=(g^{(2)}(r)-1)\langle n(r) \rangle^2$,
are shown on the bottom row of Fig.\ref{dens}.
By comparing also to HF \footnote{While the chemical potential is fixed between SGPE and modified Popov calculations,
we find a different value of $\mu$ must be used in order to match the Hartree-Fock density to
the experimental central density. The temperature however remains fixed between all methods.}, 
the trend we observe is that despite matching the experimental density 
profiles well, the HF results consistently predict density fluctuations that are too large, relative 
to those observed experimentally. This may be understood since energy reducing correlations which lead 
to the onset of quasi-condensation are not incorporated in this theory, in accordance with the 
findings for highly elongated gases presented in Ref.~\cite{Trebbia2006}.
In contrast, the SGPE and MP data agree very well with the experimental findings, 
thereby confirming the utility of each in describing the physics of 2D Bose gases 
at finite temperatures. Moreover, these findings highlight that 
choosing an optimum grid spacing based upon a higher order field correlation such 
as $g^{(2)}$ is more robust, since the density is a less sensitive measure of the system.

An important feature of Ref.~\cite{Hung2011} was the experimental demonstration of scale invariance in a 2D Bose gas.
Plotting their data against a scaled chemical potential, $\beta(\mu-V(r))$,
Hung {\it et al.} showed that densities measured at several temperatures collapsed to a single curve when appropriately 
scaled to the thermal de Broglie wavelength, $\lambda_{\rm dB}$. This was found to be true in both the thermal
and superfluid regimes; 
the fluctuation region may be identified with the crossover from a thermal gas
to a superfluid, where the system moves from an enhancement to a suppression of 
density fluctuations respectively, as may be seen in the lower rightmost plot of Figure~\ref{dens}.
\begin{figure}[t!]
  \centering
  \includegraphics[angle=0,scale=0.25,clip]{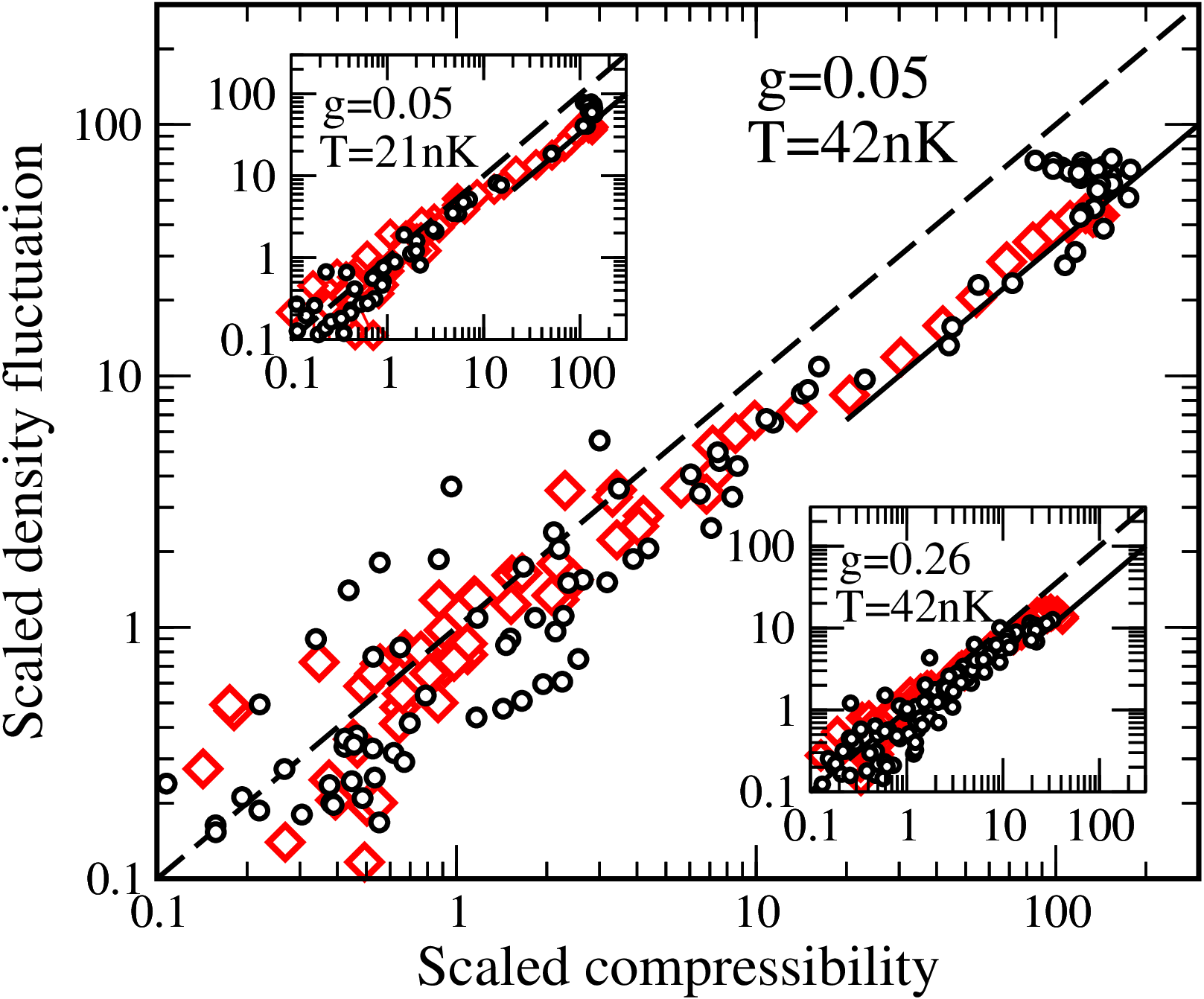}
  \caption{Scaled density fluctuation, $\langle \delta \tilde{n}^2 \rangle=\lambda_{\rm dB}^{4}\langle \delta n^2 \rangle$,
					versus the scaled compressibility, $\tilde{\kappa}=\lambda^{2}\kappa/\beta$
          from SGPE (black circles) and experiment (red diamonds). 
					Dashed and solid lines show $\langle \delta \tilde{n}^2 \rangle=\tilde{\kappa}$
					and $\langle \delta \tilde{n}^2 \rangle=\tilde{\kappa}/3$ respectively.             
          }
	\label{corr}
\end{figure}
An important difference between the MP and SGPE approaches, therefore, 
is the discontinuity in the density predicted by the former.
In contrast, a key strength of the SGPE lies in its applicability across this fluctuation region, 
as may be seen in the rightmost plots of Figure~\ref{dens}, which show 
that it also demonstrates scale invariance in good agreement with the experimental data.

To estimate the location of the BKT transition from the SGPE simulations, we employ the fitting 
function used by Hung {\it et al.}, which yields a critical chemical potential, $\mu_{\rm c}$,
and a corresponding critical density, $n_{\rm c}$.
These parameters can then be used to uncover the universal physics of 2D Bose gases,
since close to the transition point, the shifted density $(n-n_{\rm c})\lambda_{\rm dB}^{2}$ should be a 
universal function of $(\mu-\mu_{\rm c})\beta/g$ alone \cite{Prokofev2002,Hung2011}.
The universality of 2D Bose gases may then be tested by comparing scaled densities from gases with 
different interaction strengths, as in Ref.\cite{Hung2011}. As shown in Figure~\ref{crit}, by doing 
so we find that the SGPE reproduces the expected universal behaviour demonstrated experimentally. 
Moreover, we compare to the classical field MC results of Ref.~\cite{Prokofev2002},
again finding good agreement.

Finally, to examine more closely how well the experimentally observed correlation effects are captured 
by the SGPE, we compare the density fluctuations to the compressibility. The compressibility 
may be calculated from the experimental data and SGPE using the expression
$\kappa={\partial \langle n \rangle}/{\partial \mu_{\rm eff}}$, 
the results of which are shown in Figure~\ref{corr}, which once again 
closely follow the results of Ref.~\cite{Hung2011}

\textit{Conclusions --- }
We have found excellent agreement between {\it ab initio} calculations based on
the stochastic Gross-Pitaevskii and modified Popov theories
with {\it in situ} experimental data from the two-dimensional Bose gas
experiments of Hung {\it et al.}.
Combining these two methods, we devised and tested a systematic approach 
to circumvent the problem of cutoff choice within classical field simulations
that could prove crucial in modelling out-of-equilibrium scenarios,
which forms a natural extension to this work.

\textit{Acknowledgements --- } 
We thank Chen-Lung Hung and the other members of the Cheng Chin group
for supplying us with their data and for useful discussions. We acknowledge 
funding from EPSRC through grant no. EP/F055935/1.

\bibliographystyle{apsrev4-1}
\bibliography{twa-sgpe2}

\end{document}